# Symbolic Solution for Generalized Quantum Cylindrical Wells using Computer Algebra

**Edward Yesid Villegas Pulgarin**
Logic and Computation Group, Department of Physical Engineering, EAFIT University
Medellin, Antioquia, Colombia
evillega@eafit.edu.co

**Abstract –** *This paper present how to solve the problem of cylindrical quantum wells with potential energy different from zero and with singularity of the energy on the axis of the cylinder. The solution to the problem was obtained using methods of computer algebra. The results depend of Bessel and Kummer functions. This paper present energy levels and wave functions in some of the cases with an exactly form and in other cases with an approximated form, this form depended on the possibility of integrating the special functions and calculating the zeros of these functions. Here we can see the power of the method in the applications concerning complex problems of quantum mechanics, and the possibility of being able to apply this method in order to solve other problems in science and also in engineering.*

## 1 Introduction

With the big technological developments that have not only impelled to the industry but also have impelled to science and investigation, it was possible to arrive to the development of the computers and later on, to the production of a series of similar tools for these, the software [2] and applications.

These have represented in the field science and engineering a great ally when it comes to establish solutions and approaches that for an individual it would be a task to which he should dedicate a quite considerable time, being susceptible of mistakes.

Not just because of this, but also because of its versatility, it becomes necessary to implement software of computer algebra, like a primordial tool of work for the engineer and the science man. In this case it is thought to solve the problem of potential wells in cylinders with potential energy different from zero and singular in the axis, using computer algebra.

In the quantum mechanics, the problems of particles confined in potential wells are typical problems that equally possess interesting applications in the study of the decline of nuclei, the behavior of the electrons in the atom, and some other applications.

A variety of problems of this type exists, in which we can find the potential energy in a certain way or to find that this potential energy has a value similar to zero. For our final case we have the typical one-dimensional problem (a line) [1], two-dimensional (a rectangle) and three-dimensional (a cube), figure 1, whose solutions are simple and it is possible to generalize the solution. The results obtained for this problem are annexed in the table 1, 2 and 3.

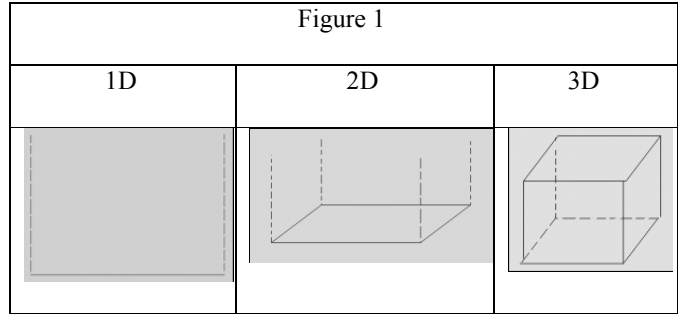

| Figure 1 | | |
|---|---|---|
| 1D | 2D | 3D |

Table 1: Solution to the problem one-dimensional

| | |
|---|---|
| Schrödinger Equation | $E\psi(x) = -\dfrac{1}{2}\dfrac{\hbar^2}{m}\dfrac{d^2\psi(x)}{dx^2}$ |
| Boundary Conditions | $\psi(0) = 0 \qquad \psi(a) = 0$ |
| Wave Function | $\psi(x) = \sqrt{\dfrac{2}{a}} \sin\left(\dfrac{n\pi x}{a}\right)$ |
| Energy Levels | $E_n = \dfrac{n^2 \pi^2 \hbar^2}{2ma^2}$ |

### Table 2: Solution to the problem two-dimensional

| | |
|---|---|
| Schrödinger Equation | $E\psi(x,y) = -\dfrac{1}{2}\dfrac{\hbar^2}{m}\left(\dfrac{\partial^2\psi(x,y)}{\partial x^2} + \dfrac{\partial^2\psi(x,y)}{\partial y^2}\right)$ |
| Boundary Conditions | $\psi(0,y)=0 \quad \psi(a,y)=0$ <br> $\psi(x,0)=0 \quad \psi(x,b)=0$ |
| Wave Function | $\psi(x,y) = \dfrac{2}{\sqrt{ab}}\sin\left(\dfrac{n\pi x}{a}\right)\sin\left(\dfrac{k\pi y}{b}\right)$ |
| Energy Levels | $E_{n,k} = \dfrac{\pi^2\hbar^2}{2m}\dfrac{(k^2 a^2 + n^2 b^2)}{a^2 b^2}$ |

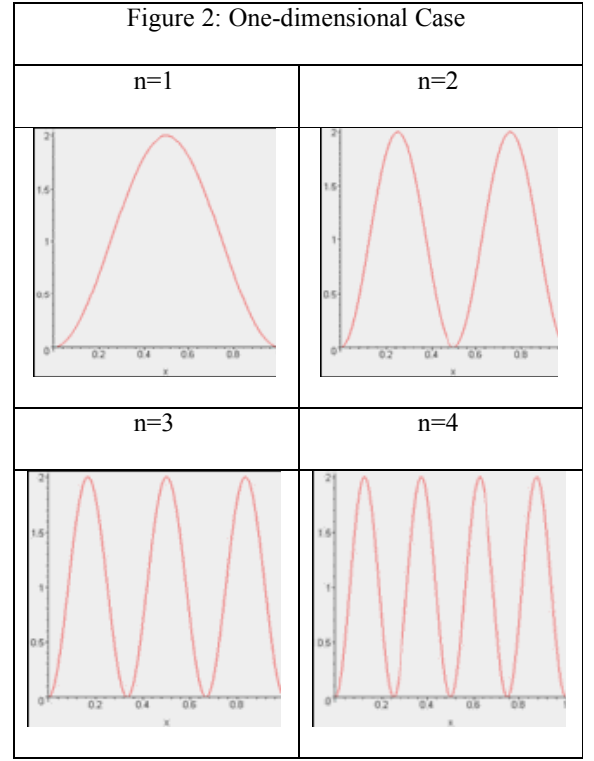

Figure 2: One-dimensional Case

| n=1 | n=2 |
|---|---|
| n=3 | n=4 |

### Table 3: Solution to the problem three-dimensional

| | |
|---|---|
| Schrödinger Equation | $E\psi(x,y,z) = -\dfrac{1}{2}\dfrac{\hbar^2}{m}\left(\dfrac{\partial^2\psi(x,y,z)}{\partial x^2} + \dfrac{\partial^2\psi(x,y,z)}{\partial y^2} + \dfrac{\partial^2\psi(x,y,z)}{\partial z^2}\right)$ |
| Boundary Condition | $\psi(0,y,z)=0 \quad \psi(x,0,z)=0 \quad \psi(x,y,0)=0$ <br> $\psi(a,y,z)=0 \quad \psi(x,b,z)=0 \quad \psi(x,y,c)=0$ |
| Wave Function | $\psi(x,y,z) = \dfrac{2\sqrt{2}}{\sqrt{abc}}\sin\left(\dfrac{n\pi x}{a}\right)\sin\left(\dfrac{k\pi y}{b}\right)\sin\left(\dfrac{N\pi y}{c}\right)$ |
| Energy Levels | $E_{n,k,N} = \dfrac{\pi^2\hbar^2}{2m}\dfrac{(N^2 a^2 b^2 + n^2 b^2 c^2 + k^2 a^2 c^2)}{a^2 b^2 c^2}$ |

Where $n$, $k$ and $N$ are integer constants; $a$, $b$ and $c$ are the values of length on the different dimensions.

We can affirm that

$$\psi(x_1,\ldots,x_n) = \sqrt{2^n}\prod_{i=1}^{n}\dfrac{\sin\left(\dfrac{N_i \pi x_i}{a_i}\right)}{\sqrt{a_i}} \quad (1)$$

And

$$E_{N_1,\ldots,N_2} = \dfrac{\pi^2\hbar^2}{2m}\sum_{i=1}^{n}\left(\dfrac{N_i}{a_i}\right)^2 \quad (2)$$

In equations (1) and (2) $n$ is the number of dimensions.

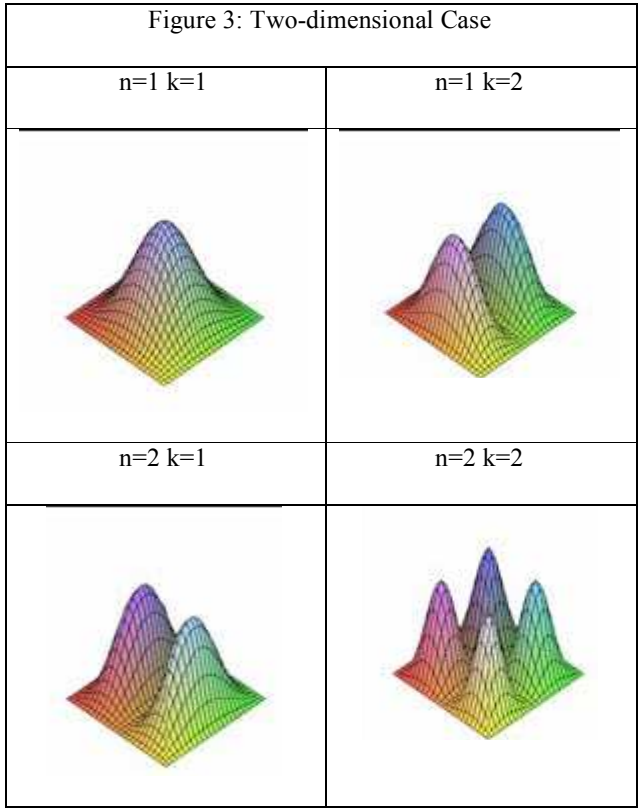

Figure 3: Two-dimensional Case

| n=1 k=1 | n=1 k=2 |
|---|---|
| n=2 k=1 | n=2 k=2 |

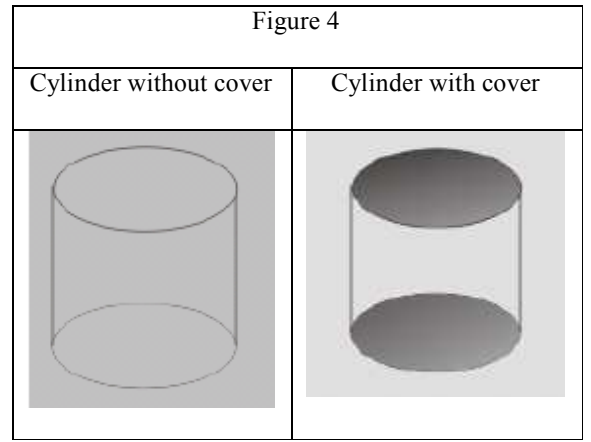

Figure 4

| Cylinder without cover | Cylinder with cover |
|---|---|

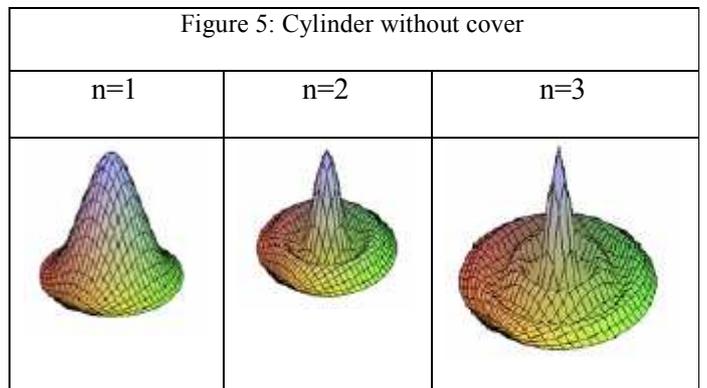

Figure 5: Cylinder without cover

| n=1 | n=2 | n=3 |
|---|---|---|

In solutions for cylindrical cases appear special functions, there are Bessel functions [3]. This might complicate the calculations a little if they are carried out in a manual way.

Table 4: Solution for cylindrical case without cover

| | |
|---|---|
| Schrödinger Equation | $E\psi(r) = -\dfrac{\hbar^2}{2m}\left(\dfrac{\partial^2 \psi(r)}{\partial r^2} + \dfrac{\partial \psi(r)}{r \partial r}\right)$ |
| Boundary Condition | $\psi(R,0) = 0$ |
| Wave Function | $\psi(r,z) = \dfrac{J_0\!\left(\dfrac{\alpha_{0,n} r}{R}\right)}{R J_1(\alpha_{0,n})\sqrt{\pi}}$ |
| Energy Levels | $E_n = \dfrac{\hbar^2}{2m}\dfrac{\alpha_{0,n}^2}{R^2}$ |

Where $n$ is a integer constant, $\alpha_{0,n}$ is $-n$ zero of $J_0(r)$, $l$ is the length of the cylinder and $R$ is the radius.

| | Table 5: Solution for cylindrical case with cover |
|---|---|
| Schrödinger Equation | $E\psi(r,z) = -\dfrac{\hbar^2}{2m}\left(\dfrac{\partial^2\psi(r,z)}{\partial r^2} + \dfrac{\partial\psi(r,z)}{r\partial r} + \dfrac{\partial^2\psi(r,z)}{\partial z^2}\right)$ |
| Boundary Condition | $\psi(r,0) = 0 \quad \psi(R,z) = 0$ <br> $\psi(r,L) = 0$ |
| Wave Function | $\psi(r,z) = \dfrac{\sqrt{2}J_0\left(\dfrac{\alpha_{0,n}r}{R}\right)\sin\left(\dfrac{n\pi z}{l}\right)}{RJ_1(\alpha_{0,n})\sqrt{l\pi}}$ |
| Energy Levels | $E_n = \dfrac{\hbar}{2m}\dfrac{n^2\pi^2 R^2 + \alpha_{0,n}^2 l^2}{l^2 R^2}$ |

## 2  The Problem

The problem that this paper presents the solution is the problem of cylindrical potential regions with singularity in the potential energy on the axis of the cylinder.

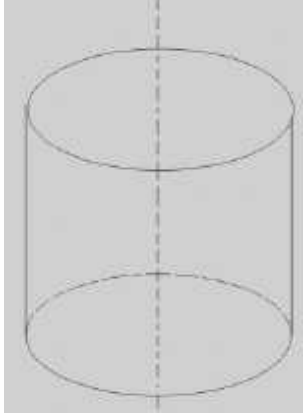

Figure 6

$$\lim_{r\to 0} U(r) = \infty$$

$U(r)$ has the form in particular

$$U(r) = \dfrac{\alpha}{r^n} \quad (3)$$

We will look for the solution for this problem in infinity cylinder and cylinders with cover.

This allows us to particularize Schrödinger Equation [1] for two cases that are going to be specified next.

## 2.1  Cylinder without cover

The Schrödinger Equation for this possesses the form

$$E\psi(r) = -\dfrac{\hbar^2}{2m}\left(\dfrac{d^2\psi(r)}{dr^2} + \dfrac{d\psi(r)}{rdr}\right) + U(r)\psi(r) \quad (4)$$

Initially we define their boundary conditions that it corresponds for this case to

$$\psi(R) = 0 \quad (5)$$

We will consider the condition of finitude of the wave function

$$\lim_{r\to 0}\psi(r) = finite \quad (6)$$

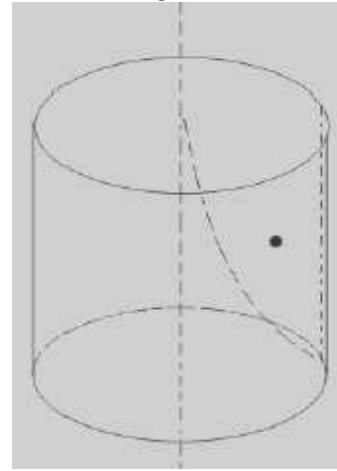

Figure 7

We will take

$$U(r) = \frac{\alpha}{r} \quad (7)$$

This way, Schrödinger Equation is expressed as

$$E\psi(r) = -\frac{\hbar^2}{2m}\left(\frac{d^2\psi(r)}{dr^2} + \frac{d\psi(r)}{rdr}\right) + \frac{\alpha}{r}\psi(r) \quad (8)$$

And

$$U(r) = \frac{\alpha}{r^2} \quad (9)$$

The Schrödinger Equation for this possesses the form

$$E\psi(r) = -\frac{\hbar^2}{2m}\left(\frac{d^2\psi(r)}{dr^2} + \frac{d\psi(r)}{rdr}\right) + \frac{\alpha}{r^2}\psi(r) \quad (10)$$

## 2.2 Cylinder with cover

In this problem the Schrödinger Equation has the general form

$$E\psi(r,z) = \left(-\frac{\hbar^2}{2m}\left(\frac{\partial^2}{\partial r^2} + \frac{\partial}{r\partial r} + \frac{\partial^2}{\partial z^2}\right) + U(r)\right)\psi(r,z) \quad (11)$$

Figure 8

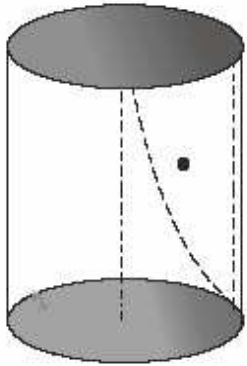

Initially we define the boundary conditions, for this case correspond to

$$\psi(r,0) = 0 \quad \psi(R,z) = 0$$
$$\psi(r,L) = 0 \quad (12)$$

And we apply the finitude condition

$$\lim_{r \to 0} \psi(r,z) = finite \quad (13)$$

We will take (7) and (9). Thus, the equation can be wrote this way as

$$E\psi(r,z) = \left(-\frac{\hbar^2}{2m}\left(\frac{\partial^2}{\partial r^2} + \frac{\partial}{r\partial r} + \frac{\partial^2}{\partial z^2}\right) + \frac{\alpha}{r}\right)\psi(r,z) \quad (14)$$

and

$$E\psi(r,z) = \left(-\frac{\hbar^2}{2m}\left(\frac{\partial^2}{\partial r^2} + \frac{\partial}{r\partial r} + \frac{\partial^2}{\partial z^2}\right) + \frac{\alpha}{r^2}\right)\psi(r,z) \quad (15)$$

respectively.

## 3 Method of Solution

Due to the complexity of the problem to solve it becomes necessary the use of computer algebra systems [6], they allows us to obtain solutions to this problem in a short lapse of time, and these solutions bring a great reliability.

When looking at the differential equation of the problem, we can see that the solution of these is not constituted by simple functions, as polynomial, exponential or trigonometric function, but rather should be identified in them some general structures for the recognition of a special function [7] this function is solution to the differential equation. Task not very simple if the differential equation possesses great quantity of algebraic objects, an issue problem for which the system of computer algebra will be of great utility.

That is the reason why a system of computer algebra [6] will be useful to achieve the solution of this problem; in this case this software is Maple [2].

## 4 Results

The obtaining of the following results was achieved with the help of software [2] of computer algebra [6].

### 4.1 Solution for a cylinder infinity (without cover)

In the first case, when the boundary condition is specified by (7), due to the impossibility of finding all roots of the Kummer function [4] in an analytical way, it is necessary to carry out an expansion of this and to approach its first roots. For this solution we approach its first two roots and we obtain the wave functions corresponding to each energy level.

$$E_1 = \frac{2\alpha^2 m}{\hbar^2} \quad (16)$$

$$E_2 = \frac{2(\hbar^2 + R\alpha m)^2}{R^2 \hbar^2 m} \quad (17)$$

$$\psi(r) = 135\sqrt{77}\, h^{10}\, e^{\left(\frac{-2 I m \alpha r}{h^2}\right)} \text{KummerM}\left(\frac{1}{2} - \frac{1}{2}I, 1, \frac{4 I m \alpha r}{h^2}\right) / ((\pi\,($$

$$4158\, m^{10} \alpha^{10} R^{10} - 7560\, m^9 \alpha^9 R^9 h^2 - 63063\, m^8 \alpha^8 R^8 h^4 + 61600\, m^7 \alpha^7 R^7 h^6 \quad (18)$$

$$+ 464310\, m^6 \alpha^6 R^6 h^8 - 71280\, m^5 \alpha^5 R^5 h^{10} - 1767150\, m^4 \alpha^4 R^4 h^{12}$$

$$- 997920\, m^3 \alpha^3 R^3 h^{14} + 2806650\, m^2 \alpha^2 R^2 h^{16} + 3742200\, R \alpha m h^{18} + 1403325\, h^{20}$$

$$))^{(1/2)} R)$$

$$\psi_2 = 3780\, h^8\, e^{\left(\frac{-2 I (h^2 + R \alpha m) r}{h^2 R}\right)} \text{KummerM}\left(-\frac{-h^2 - R\alpha m + \alpha m R I}{2(h^2 + R\alpha m)}, 1, \quad (19)\right.$$

$$\left.\frac{4 I (h^2 + R \alpha m) r}{h^2 R}\right) / (4643100\, R \alpha m \pi h^{14} + 35280\, R^8 \pi m^8 \alpha^8 + 5536755\, \pi h^{16}$$

$$+ 486080\, R^7 \pi m^7 \alpha^7 h^2 + 6037360\, R^5 \pi m^5 \alpha^5 h^6 + 4212320\, R^4 \pi h^8 m^4 \alpha^4$$

$$- 5418000\, R^2 \pi h^{12} m^2 \alpha^2 - 5209120\, R^3 \pi h^{10} m^3 \alpha^3 + 2571520\, R^6 \pi m^6 \alpha^6 h^4)^{(1/2)} R)$$

For the case two (boundary condition 8), we have that

$$E_n = \frac{1}{2}\frac{z_n^2 \hbar^2}{mR^2} \quad (20)$$

$$\psi(r) := 2^{\left(-\frac{\sqrt{2}\sqrt{\alpha}\sqrt{m}}{h}\right)} z_n^{\left(-\frac{\sqrt{2}\sqrt{\alpha}\sqrt{m}}{h}\right)} \sqrt{h}\, 4^{\left(\frac{\sqrt{2}\sqrt{\alpha}\sqrt{m}}{h}\right)} \Gamma\left(\frac{2h + \sqrt{2}\sqrt{\alpha}\sqrt{m}}{h}\right) \quad (21)$$

$$\text{BesselJ}\left(\frac{\sqrt{2}\sqrt{\alpha}\sqrt{m}}{h}, \frac{z_n r}{R}\right) \Big/ \Bigg( R$$

$$\sqrt{\text{hypergeom}\left(\left[\frac{2\sqrt{2}\sqrt{\alpha}\sqrt{m} + h}{2h}\right], \left[\frac{2h + \sqrt{2}\sqrt{\alpha}\sqrt{m}}{h}, \frac{2\sqrt{2}\sqrt{\alpha}\sqrt{m} + h}{h}\right], -z_n^2\right)}$$

$$\sqrt{\pi}\sqrt{h + \sqrt{2}\sqrt{\alpha}\sqrt{m}}\Bigg)$$

Where $z_n$ is the $-n$ zero of $J_{\frac{\sqrt{2m\alpha}}{\hbar}}(r)$.

## 4.2 Solution for a cylinder with cover

Firstly, we obtain the solution for the case with boundary condition specified in equation 7.

$$E_n = \frac{-L^2\left(\hbar^4 + 4\hbar^2 R\alpha m + 4R^2\alpha^2 m^2 L^2\right) + n^2\pi^2\hbar^4 R^2}{2mR^2\hbar^2 L^2} \quad (22)$$

and the energy levels for the second boundary condition

$$E_{n,N} = \frac{\hbar^2}{2}\frac{n^2\pi^2 R^2 + L^2 z_N^2}{mL^2 R^2} \quad (23)$$

And their wave functions respectively

$$\psi(r,z) := \frac{e^{\left(-\frac{(h^2+2R\alpha m)r}{h^2 R}\right)} \sqrt{6}\, h^2\, \text{KummerM}\left(\frac{h^2+4R\alpha m}{2(h^2+2R\alpha m)}, 1, \frac{2(h^2+2R\alpha m)r}{h^2 R}\right) \sin\left(\frac{n\pi z}{L}\right)}{\sqrt{\pi L\,(6R^2\alpha^2 m^2 + 8h^2 R\alpha m + 3h^4)\,R}}$$

(24)

And

$$\psi(r,z) = \frac{2^{\frac{\sqrt{2m\alpha}}{\hbar}+\frac{1}{2}} z_N^{\frac{-\sqrt{2m\alpha}}{\hbar}} \sqrt{\hbar+\sqrt{2m\alpha}}\, \Gamma\!\left(\frac{\hbar+\sqrt{2m\alpha}}{\hbar}\right) J_{\frac{\sqrt{2m\alpha}}{\hbar}}\!\left(\frac{z_N r}{R}\right) \sin\!\left(\frac{n\pi z}{L}\right)}{R\sqrt{\text{hypergeom}\!\left(\frac{\sqrt{2m\alpha}}{\hbar}+\frac{1}{2}, \left[\frac{2\sqrt{2m\alpha}+\hbar}{\hbar}, \frac{\sqrt{2m\alpha}+2\hbar}{\hbar}\right], -z_N^2\right) L\pi\hbar}}$$

(25)

## 5 Conclusions

It is possible to observe how the solution of the Schrödinger Equation [1] for the cases where U=0 in the interior of the region, they offer a certain easiness to find their solution manually, although for the cylindrical cases the problem requires of the use of special functions, then the difficulty or time of development of the problem begin to increase.

As we pretend to find solutions for problems every time more complicated (problems for those which U is function of space variables), the capacity and human ability to develop this type of problems begin to diminish, and this activity begins to lose effectiveness due to the extension of their solutions as for the time of its obtaining.

Another important point is also the increase of difficulty in the recognition of the solution functions (special) of Schrödinger Equation for each particular case.

It is here where the necessity of the using the software [2] of computer algebra [6] appears, to be able to speed up the process of solution of problems that could take a considerable time because its complexity and extension.

It is then when the computer algebra system [6] is a powerful tool to compute and it is also a library of special functions [7] that facilitate the solution process (in particular of symbolic type – it is in the cases that one observes their efficiency – ) to multiple quite tedious problems.